\documentclass[paper]{ieice}
\usepackage[dvipdfmx]{graphicx,xcolor}
\usepackage[fleqn]{amsmath}
\usepackage{newtxtext}
\usepackage[varg]{newtxmath}
\usepackage{subfigure}
\usepackage{cite}

\field{D}
\vol{101}
\no{12}
\title{A Block-Permutation-Based Encryption Scheme with Independent Processing of RGB Components}
\authorlist{%
 \authorentry[imaizumi@chiba-u.jp]{Shoko IMAIZUMI}{m}{cu}\MembershipNumber{0215457}
 \authorentry[kiya@sd.tmu.ac.jp]{Hitoshi KIYA}{f}{tmu}\MembershipNumber{}
}
\affiliate[cu]{The authors is with the Graduate School of Engineering,
Chiba University, Chiba-shi, 263-8522 Japan.
}
\affiliate[tmu]{The author is with the Faculty of System Design, Tokyo
Metropolitan University, Hino-shi, 191-0065 Japan.
}

\received{2018}{3}{3}
\revised{2015}{7}{26}

\begin{document}
\maketitle
\begin{summary}
 This paper proposes a block-permutation-based encryption (BPBE) scheme for the encryption-then-compression (ETC) system that enhances the color scrambling.
 A BPBE image can be obtained through four processes, positional scrambling, block rotation/flip, negative-positive transformation, and color component shuffling, after dividing the original image into multiple blocks.
 The proposed scheme scrambles the R, G, and B components independently in positional scrambling, block rotation/flip, and negative-positive transformation, by assigning different keys to each color component.
 The conventional scheme considers the compression efficiency using JPEG and JPEG 2000, which need a color conversion before the compression process by default.  
 Therefore, the conventional scheme scrambles the color components identically in each process. 
 In contrast, the proposed scheme takes into account the RGB-based compression, such as JPEG-LS, and thus can increase the extent of the scrambling.
 The resilience against jigsaw puzzle solver (JPS) can consequently be increased owing to the wider color distribution of the BPBE image. 
 Additionally, the key space for resilience against brute-force attacks has also been expanded exponentially.
 Furthermore, the proposed scheme can maintain the JPEG-LS compression efficiency compared to the conventional scheme.
 We confirm the effectiveness of the proposed scheme by experiments and analyses. 
\end{summary}

\section{Introduction}
\label{sec:1}
 Privacy and copyright protection for digital images has been a serious concern in cloud services, social networking services, and so forth. 
 The traditional technique to securely transmit images is the Compression-then-Encryption (CtE) system, which performs compression before encryption. 
 However, the image owner has to disclose the image content to a network provider in this system. 
 Therefore, another system for image transmission, that is, the Encryption-then-Compression (EtC) system, has been studied as a framework where encryption is performed by the image owner before compression/transmission~\cite{Elsevier_DSP2017:MKumar, IEEE-T2014:JZhou, IEEE-T2010:WLiu, IEEE-T2004:MJohnson}. 
 Common key cryptosystems, such as the AES and the triple DES, are frequently used for image protection. 
 However, there is a trade-off between security and additional signal processing in the encryption domain for image transmission systems. 
 For this reason, many kinds of encryption algorithms based on scrambling have been studied.

 In this paper, we focus on block-permutation-based encryption (BPBE), which is an image-scrambling encryption technique~\cite{IEICE-T2017:KKurihara, BMSB2016:KKurihara, IEICE-T2015:KKurihara, ICASSP2015:OWatanabe}.
 The conventional BPBE scheme first divides the original image into definite-sized blocks and then performs four processes. 
 The main feature of the conventional scheme is to maintain the compression efficiency of BPBE images using JPEG and JPEG 2000, which need a color conversion before the compression process by default.  
 Therefore, the conventional scheme scrambles the R, G, and B components identically in each process. 
 Consequently, the color distribution of the original image deeply affects that of its encrypted image. 
 We propose an extended BPBE algorithm to deal with this issue.
 The proposed scheme considers the RGB-based compression technique such as JPEG-LS~\cite{IEEE-T:MJWeinberger}, which is an international lossless coding standard.
 Accordingly, our scheme can scramble the three color components independently without degrading the compression efficiency and thus increases the extent of the scrambling.
 
 Jigsaw puzzle solver (JPS)~\cite{GPEM2016:DSholomon, CVPR2012:ACGallagher, CVPR2016:KSon, CVPR2015:GPaikin, ECCV2014:KSon, CVPR2011:DPomeraz, CVPR2010:TCho}, which has been studied in the computer vision and pattern recognition fields, is an attack to illegally retrieve the original image from multiple pieces by using pixel correlations.
 Because the BPBE scheme is an image-scrambling encryption technique using multiple blocks, there is still some correlation among the pixels in the encrypted image. 
 We take into consideration the security against JPS. 
 We confirm that it is difficult to retrieve the original image by JPS in the proposed scheme.
  
 Owing to the proposed algorithm that assigns different keys to each color component, the key space for resilience against brute-force attacks can be expanded exponentially. 
 On the other hand, the compression efficiency of the proposed scheme using JPEG-LS is almost the same as that of the conventional scheme. 
 Our experimental results show the effectiveness of the proposed scheme.

\section{Preparation}
\label{sec:2}

\subsection{Block-permutation-based encryption}
\label{ssec:2-1}
\begin{figure}[tb]
 \centering

  \includegraphics[width=0.988\columnwidth]{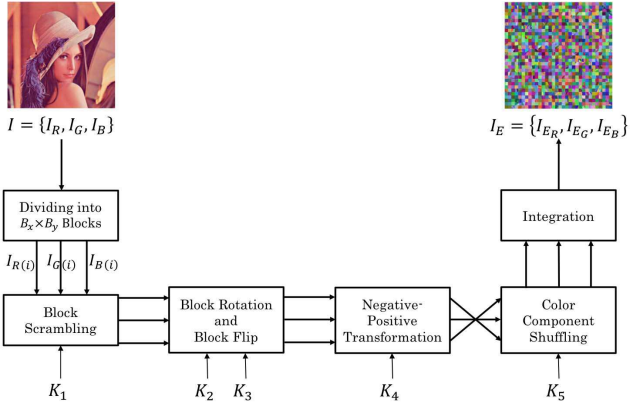}
 \caption{Procedure of block-permutation-based encryption.}
 \label{fig:bpbe}
\end{figure}

 We first describe the fundamental procedure of the conventional BPBE scheme~\cite{IEICE-T2017:KKurihara}. 
 The image owner does not need to disclose the image content to a network provider by encrypting the original image before sending it to the provider. 
 As shown in Fig.~\ref{fig:bpbe}, the BPBE scheme first divides the original image into definite-sized blocks and then executes four processes: positional scrambling, block rotation/flip, negative-positive transformation, and color component shuffling. 
 Finally, it integrates the blocks into one encrypted image. 
 The BPBE scheme can control the quality of the encrypted image and the encryption strength by changing the block size. 
 In addition, the compression efficiency of the encrypted image becomes equivalent to that of the original image. 
 The encryption procedure is described in what follows.

\begin{description}
 \item{\bf{Step 1}:} Divide the original image $I =\{I_R, I_G, I_B\}$ with $M \times N$ pixels into multiple blocks with $B_x \times B_y$ pixels.
 \item{\bf{Step 2}:} Scramble the position of each block using a random number generated by key $K_1$.
 \item{\bf{Step 3}:} Rotate and flip each block using random numbers generated by keys $K_2$ and $K_3$.
 \item{\bf{Step 4}:} Perform negative-positive transformation on each block using a random number generated by key $K_4$.
 \item{\bf{Step 5}:} Shuffle the R, G, and B components in each block using a random number generated by key $K_5$.
 \item{\bf{Step 6}:} Integrate all blocks and generate the encrypted image.
\end{description}

 Note that keys $K_1$, $K_2$, $K_3$, and $K_4$ are commonly used for the three color components in the conventional scheme.
 The conventional scheme would not only change the spatial positions and directions but would also reverse the pixel values and permute the three color components in each block. 
 However, the color distribution of the original image directly affects that of the encrypted image. 
 We propose a new algorithm to reduce the effect of the color distribution of the original image and increase the extent of the scrambling in Section~\ref{sec:3}.

\subsection{Jigsaw puzzle solving problems}
\label{ssec:2-2}
 JPS is an attack that tries to retrieve the original image from multiple pieces by utilizing the correlation among them. 
 JPS has been studied in the computer vision and pattern recognition fields. 
 The safety of the BPBE schemes has been discussed mainly with regards to the key spaces for resilience against brute-force attacks. 
 However, because the BPBE images consist of multiple blocks, JPS should also be considered as one of the attacks on the BPBE schemes. 
 It has been reported that a jigsaw puzzle consisting of 30,745 pieces can be solved completely using the conventional JPS~\cite{GPEM2016:DSholomon}. 
 Another JPS has succeeded in solving a puzzle where the directional information of each piece is not apparent~\cite{CVPR2012:ACGallagher}. 
 It has been confirmed that some encryption schemes, where the key spaces are sufficiently large, are still vulnerable against JPS. 
 On the other hand, it is difficult for the conventional JPSs to solve a puzzle where the color distribution is modified~\cite{IEICE-T2018:TChuman, ICASSP2017:TChuman, ICME2017:TChuman, ICME2018:WSirichotedumrong}.  

 Security against other attacking strategies, such as know-plaintext attack (KPA) and chosen-plaintext attack (CPA), is frequently discussed in addition to JPS. 
 First, BPBE prepares different encryption keys to each image and thus is robust against KPA. 
 Secondly, BPBE is not a public key encryption scheme.
 This means that the encryption keys for BPBE do not need to be disclosed and can be kept confidential. 
 CPA is necessarily prevented in BPBE in contrast to public key encryption schemes.

 In the next section, we propose a new BPBE approach to enhance the color scrambling.

\section{Proposed scheme}
\label{sec:3}
 We propose a BPBE algorithm to increase the extent of the scrambling for the security against JPS. 
 The conventional scheme aims to maintain the compression efficiency using JPEG and JPEG 2000, which need color conversion before the compression process by default.
 Therefore, it needs to scramble the three color components identically using a single key in each process. 
 On the other hand, the proposed scheme takes into account the RGB-based compression such as JPEG-LS and thus can deal with the color components independently.
 Consequently, the color distribution of the original image does not severely affect that of its encrypted image.
 The compression efficiency using JPEG-LS is equivalent to that of the conventional scheme.
 The proposed BPBE procedure is described by the following steps. 
 
 \begin{description}
 \item{\bf{Step 1}:} Divide the original image $I =\{I_R, I_G, I_B\}$ with $M \times N$ pixels into multiple blocks with $B_x \times B_y$ pixels.
 \item{\bf{Step 2}:} Scramble the position of each block using random numbers generated by three keys: $K_{1,R}$, $K_{1,G}$, and $K_{1,B}$.
 \item{\bf{Step 3}:} Rotate and flip each block using random numbers generated by six keys: $K_{2,R}$, $K_{2,G}$, $K_{2,B}$, $K_{3,R}$, $K_{3,G}$, and $K_{3,B}$.
 \item{\bf{Step 4}:} Perform the negative-positive transformation on each block using random numbers generated by three keys: $K_{4,R}$, $K_{4,G}$, and $K_{4,B}$.
 \item{\bf{Step 5}:} Shuffle the R, G, and B components in each block using a random number generated by key $K_5$.
 \item{\bf{Step 6}:} Integrate all blocks and generate the encrypted image.
\end{description}

 Note that $K_{\alpha,R}$, $K_{\alpha,G}$, and $K_{\alpha,B}$ $(\alpha= 1, 2, 3, 4)$ are the keys for the R, G, and B components, respectively. 
 Therefore, each color component can be independently scrambled using different keys.
  
\subsection{Individual processing among RGB color components}
\label{ssec:3-1}
\begin{figure}[tb]
 \centering

  \includegraphics[width=0.44\columnwidth]{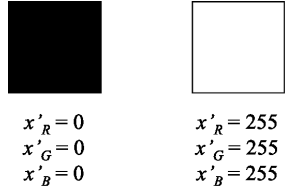}
 \caption{Example of negative-positive transformation when using conventional scheme~\cite{IEICE-T2017:KKurihara} where $\{x_R, x_G, x_B\} = \{255, 255, 255\}$.}
 \label{fig:negaposi_conv}
\end{figure}

\begin{figure}[tb]
 \centering

  \includegraphics[width=0.95\columnwidth]{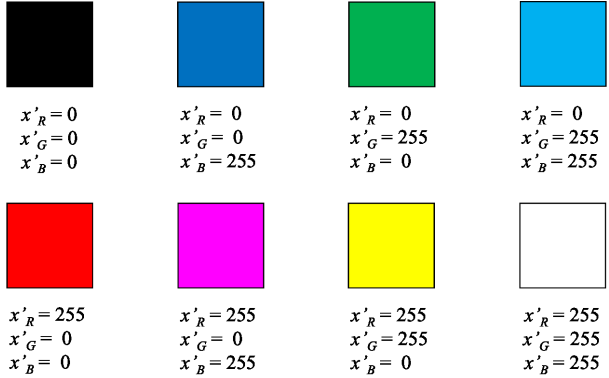}
 \caption{Example of negative-positive transformation when using proposed scheme where $\{x_R, x_G, x_B\} = \{255, 255, 255\}$.}
 \label{fig:negaposi_prop}
\end{figure}

 As described in \ref{ssec:2-1}, keys $K_1$, $K_2$, $K_3$, and $K_4$ for the three processes, positional scrambling, block rotation/flip, and negative-positive transformation, are commonly used for the R, G, and B components in the conventional scheme~\cite{IEICE-T2017:KKurihara}. 
 This means that the three color components in each block are identically scrambled in the three processes.
 The proposed scheme prepares three keys for each process, e.g., $K_{1,R}$, $K_{1,G}$, and $K_{1,B}$ for the positional scrambling, and independently scrambles the three color components. We give a detailed account using the negative-positive transformation as follows.
 
 In the conventional scheme, the negative-positive transformation is identically operated for the three color components by using a random number of either zero or one, which is generated by key $K_4$. 
 In the case that pixel $x$ has values $\{x_R, x_G, x_B\} = \{255, 255, 255\}$, they would be changed to $\{x'_R, x'_G, x'_B\} = \{0, 0, 0\}$ or would not be changed, that is, $\{x'_R, x'_G, x'_B\} = \{255, 255, 255\}$, as shown in Fig.~\ref{fig:negaposi_conv}. 
 On the other hand, the proposed scheme prepares three random numbers of zero or one, which are generated by three keys $K_{4,R}$, $K_{4,G}$, and $K_{4,B}$. 
 The negative-positive transformation of the three color components are operated independently according to the random numbers. 
 In the case of the above example where $\{x_R, x_G, x_B\} = \{255, 255, 255\}$, they could be changed to $\{x'_R, x'_G, x'_B\} = \{0, 0, 0\}$, $\{0, 0, 255\}$, $\{0, 255, 0\}$, \{0, 255, 255\}, $\{255, 0, 0\}$, $\{255, 0, 255\}$, $\{255, 255, 0\}$, or $\{255, 255, 255\}$, as shown in Fig.~\ref{fig:negaposi_prop}.
 Consequently, the proposed scheme can increase the extent of the scrambling by using independent keys for the R, G, and B components. 
 Here, we generalize the above example.
 Each color component of pixel $x$ is represented by $x_c ~(c\in\{R, G, B\})$, and is independently transformed by using key $K_{4,c}$.
 The color components after the negative-positive transformation are given by
 \begin{eqnarray}
  x'_c=\left\{ \begin{array}{ll}
              x_c, &  \textrm{if} ~~ r(x_c) =0 \\
              255-x_c, & \textrm{if} ~~ r(x_c) =1, \\
             \end{array} \right.
 \label{eq:x'_c}
 \end{eqnarray}
 where $r(x_c)$ is a random integer given for $x_c$ by using $K_{4,c}$.

\subsection{Analysis of key space for resilience against brute-force attacks}
\label{ssec:3-2}
 We discuss the safety of the proposed algorithm with its key space for the security against brute-force attacks here. 
 The four encryption processes are independent from each other. 
 Thus, the total key space can be obtained by multiplying the key spaces for the four processes.
 In the case of dividing an $M \times N$ image into $B_x \times B_y$ blocks, the number of divided blocks $L$ is given as
 \begin{equation}
  L = \lfloor \frac{M}{B_x} \rfloor \times \lfloor \frac{N}{B_y} \rfloor. \label{eq:L}
 \end{equation}
 In the positional scrambling, key space $N_P$, which is the number of all the scrambling patterns of $L$ blocks, is calculated by
 \begin{equation}
  N_{P} = ({}_L P_L)^3 = (L!)^3. \label{eq:N_P}
 \end{equation}
 The numbers of all patterns for both the block rotation and the block flip are four and four, respectively. 
 When combining those two processes, some combinations correspond to other combinations. 
 Therefore, the number of total patterns for the block rotation/flip becomes eight. 
 Combined key space of the block rotation and flip $N_{D}$ is computed by
 \begin{equation}
  N_{D} = (8^L)^3 = 512^L. \label{eq:N_D}
 \end{equation}
 The numbers of all patterns on the negative-positive transformation and the color component shuffling for each block are eight and six, respectively. 
 Key spaces $N_{N}$ and $N_{C}$ are given by
 \begin{align}
  N_{N} &= (2^L)^3 = 8^L, \\ 
  N_{C} &= 6^L. \label{eq:N_C}
 \end{align}

 Consequently, total key space $N_{A}$ in the proposed scheme can be represented by
 \begin{align}
  N_{A} &= N_{P} \times N_{D} \times N_{N} \times N_{C} \notag \\
        &= (L!)^3 \times 512^L \times 8^L \times 6^L, \label{eq:N_A}
 \end{align}
 while total key space in the conventional
 scheme~\cite{IEICE-T2017:KKurihara} $N_{A, Conv}$ is given as
 \begin{align}
  N_{A, Conv} &= N_{P, Conv} \times N_{D, Conv} \times N_{N, Conv}
  \times N_{C, Conv} \notag \\
        &= L! \times 8^L \times 2^L \times 6^L, \label{eq:N_A}
 \end{align}
 where $N_{P, Conv}$, $N_{D, Conv}$, $N_{N, Conv}$, and $N_{C, Conv}$ are the key spaces for the above four encryption processes in the conventional scheme.
 
 Accordingly, it is evident that the resilience against brute-force attacks in the proposed scheme has been significantly improved compared to that of the conventional scheme.

\section{Experimental results}
\label{sec:4}
 We evaluate the effectiveness of the proposed BPBE algorithm from the aspects of JPS resilience, color distribution, and compression efficiency using JPEG-LS. 
 The seven $512 \times 512$ images, that is, Airplane, Tiffany, Lena, Mandrill, Peppers, Sailboat, and Splash, were used as test images.

\subsection{JPS resilience}
\label{ssec:4-1}
 As described in \ref{ssec:2-2}, JPS could be considered as one of the possible attacks on the BPBE schemes because a BPBE image consists of multiple blocks. 
 The existing JPSs are broadly classified into three approaches depending on the assembly strategies: greedy algorithms, heuristic global algorithms, and linear programming algorithm\cite{arXiv:RYu}. 
 The greedy algorithms start from the initial pairwise matches and sequentially assemble the larger components but are sensitive local minima.
 The heuristic global algorithms directly search for a solution by maximizing a global compatibility function.
 However, they can be interrupted by a combinatorial search over the placement of ambiguous pieces.
 The above two algorithms are integrated into the linear programming algorithm with both the reduced sensitivity to local minima and the increased robustness to the presence of mismatches in the pairwise matches.
 They all support for puzzles with unknown positional scrambling and unknown rotation.  
 In contrast, puzzles including unknown flip, unknown negative-positive transformation, or unknown color component shuffling are not supported in the existing JPSs.

 It has been demonstrated that the restorability by JPS can be decreased a great deal in the case that the color information of the encrypted image has been modified~\cite{IEICE-T2018:TChuman, ICASSP2017:TChuman, ICME2017:TChuman, ICME2018:WSirichotedumrong}. 
 It would be difficult for the conventional JPSs to solve the puzzle when the color correlation among the pieces is low.   
 Therefore, the proposed algorithm, which modifies the color distribution of the encrypted image, can be effective against JPS. 
 We prove this below. 

 \subsubsection{Conditions}
 \label{sssec:4-1-1}
 The $512 \times 512$ images have previously been clipped to $512 \times 480$ pixels due to the analysis of the JPS resilience. 
 The JPS algorithms cannot retrieve any original images in the case that the original image is square. 
 Therefore, we use the $512 \times 480$ clipped images for the evaluation of the JPS resilience.
 The divided block size is $32 \times 32$ pixels.

 We compare assembled image $I_d$ with its original image. 
 The three types of evaluation criteria \cite{CVPR2012:ACGallagher, CVPR2010:TCho} are introduced in our experiment.
 \begin{itemize}
 \item {\bf{Direct comparison ($D_c$)}} is the ratio of the number of pieces that have been allocated in the correct position.
  $D_c$ for assembled image $I_d$, which is represented as $D_c(I_d)$, is given by
  \begin{align}
   D_c(I_d) &= \frac{1}{n}\sum^{n}_{i=1}d_c(i), \notag \\
   d_c(i) &=  \begin{cases}
    1, &\rm{if}~\it{I_d(i)}~\rm{is~allocated~in~the~correct}\\
       &\rm{position} \\
    0, &\rm{otherwise}, 
    \end{cases} \label{eq:DcId}
  \end{align}
 where $I_d(i)$ represents the position of piece $i$ in assembled image $I_d$.
 \item {\bf{Neighbor comparison ($N_c$)}} is the ratio of the number of pairwise-block adjacencies that have been correctly concatenated.  
 $N_c$ for assembled image $I_d$, which is represented as $N_c(I_d)$, is obtained by
  \begin{align}
   N_c(I_d) &= \frac{1}{B}\sum^{B}_{k=1}n_c(k), \notag \\
   n_c(k) &=  \begin{cases}
    1, &\rm{if}~\it{b_k}~\rm{is~concatenated~correctly}\\
    0, &\rm{otherwise}, 
    \end{cases} \label{eq:NcId}
  \end{align}
  where $B$ represents the number of block adjacencies in assembled image $I_d$, and $b_k$ is the $k$-th block adjacency. 
  In the case that a target image has $u \times v$ blocks, the total number of block adjacencies in $I_d$ becomes $2uv - u - v$.
 \item {\bf{Largest component ($L_c$)}} is the ratio of the largest number of concatenated blocks in a correctly assembled region.
 $L_c$ for assembled image $I_d$, which is represented as $L_c(I_d)$, is calculated by
  \begin{equation}
   L_c(I_d) = \frac{1}{n}\max_{j} \{I_c(I_d,j)\}, \quad j = 1,2, \cdots, m, \label{eq:LcId} \\
  \end{equation}
 where $I_c(I_d,j)$ represents the number of blocks in the $j$-th correctly assembled region,
 and $m$ is the number of correctly assembled regions.
 \end{itemize}
 The ranges of $D_c(I_d)$, $N_c(I_d)$, and $L_c(I_d)$ are 0 to 1, namely, $D_c(I_d), N_c(I_d), L_c(I_d) \in [0,1]$.
 The larger each value of the evaluation criteria becomes, the higher the compatibility of JPS is. 

 We produce ten encrypted images for each test image using different encryption keys. 
 The assembled image produced by JPS, which has the highest score of $D_c(I_d) + N_c(I_d) + L_c(I_d)$ in the ten images, is adopted for the following evaluation. 
 
 \subsubsection{Evaluation}
 \label{sssec:4-1-2}
 \begin{figure}[t]
 \centering

  \subfigure[Original images]{%
    \includegraphics[width=.8\columnwidth]{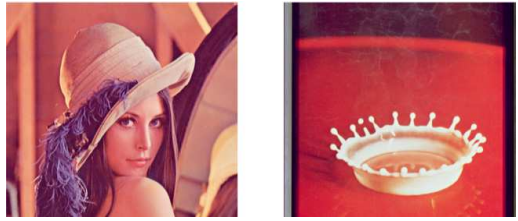}%
    \label{sfig:ori_jps}%
   }%

  \subfigure[Assembled images produced by JPS in proposed scheme (Lena: $L_c(I_d)$=0.0083, Splash: $L_c(I_d)$=0.0083)]{%
   \includegraphics[width=.8\columnwidth]{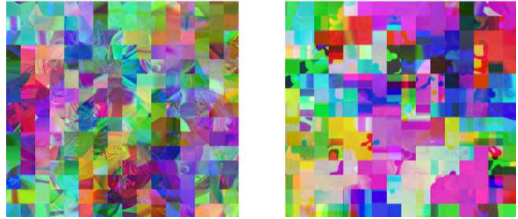}%
   \label{sfig:indep_jps}%
  }%

  \subfigure[Assembled images produced by JPS in conventional scheme (Lena: $L_c(I_d)$=0.2167, Splash: $L_c(I_d)$=0.4958)]{%
   \includegraphics[width=.8\columnwidth]{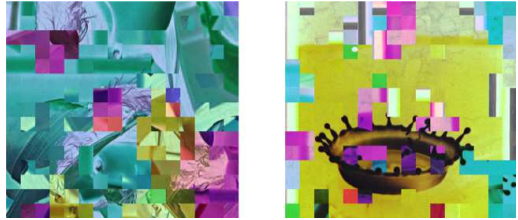}%
   \label{sfig:comm_jps}%
  }%
 \caption{Assembled images produced by JPS (Airplane and Splash).}
 \label{fig:assembled_images}
 \end{figure} 

 \begin{table*}[t]
  \begin{center}
  \caption{Ratio of correct block assembling by JPS.}
  \label{tb:jps}
   \begin{tabular}{|c|c|c|c|c|c|c|} \hline
      & \multicolumn{3}{|c|}{Prop.} & \multicolumn{3}{|c|}{Conv.} \\ \cline{2-7}
    Image & $D_c(I_d)$ & $N_c(I_d)$ & $L_c(I_d)$ & $D_c(I_d)$ & $N_c(I_d)$ & $L_c(I_d)$ \\ \hline\hline
    Airplane & 0.0000 & 0.0045 & 0.0083 & 0.0042 & 0.2149 & 0.2750 \\ \hline
    Tiffany & 0.0167 & 0.0078 & 0.0125 & 0.0000 & 0.1258 & 0.1667 \\ \hline
    Lena & 0.0000 & 0.0056 & 0.0083 & 0.2167 & 0.1526 & 0.2167 \\ \hline
    Mandrill & 0.0042 & 0.0056 & 0.0042 & 0.0000 & 0.0724 & 0.0666\\ \hline
    Peppers & 0.0000 & 0.0056 & 0.0083 & 0.0000 & 0.0445 & 0.0625 \\ \hline
    Sailboat & 0.0000 & 0.0056 & 0.0083 & 0.0000 & 0.0768 & 0.0958 \\ \hline
    Splash & 0.0042 & 0.0056 & 0.0083 & 0.4958 & 0.3820 & 0.4958 \\ \hline\hline
    \bf{Average} & \bf{0.0036} & \bf{0.0057} & \bf{0.0083} & \bf{0.1024} & \bf{0.1527} & \bf{0.1970} \\ \hline
   \end{tabular}
  \end{center}
 \end{table*}

 Fig.~\ref{fig:assembled_images} shows the assembled images for two test images in the proposed and the conventional schemes. 
 The conventional encrypted images were correctly assembled in multiple regions, and some outlines of the original images have been exposed.
 On the other hand, the assembled images produced by JPS in proposed scheme hardly reveal their image content.
 Table~\ref{tb:jps} demonstrates the concrete values of $D_c(I_d)$, $N_c(I_d)$, and $L_c(I_d)$.
 The average value of $D_c(I_d) + N_c(I_d) + L_c(I_d)$ in the proposed scheme is less than 1/25 of that in the conventional scheme.
 Consequently, it is proved that the proposed scheme has more strong resilience against JPS than the conventional scheme.

\subsection{Color distribution}
\label{ssec:4-2}
 \begin{figure}[t]
  \centering

  \subfigure [Original images (Airplane: $H(A)$=13.96, Tiffany: $H(A)$=14.66)]{%
    \includegraphics[width=.8\columnwidth]{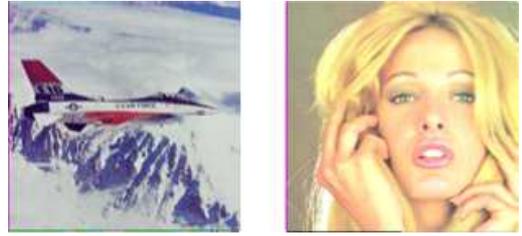}%
    \label{sfig:originals}%
   }%

  \subfigure[Encrypted images using proposed scheme (Airplane: $H(A)$=17.76, Tiffany: $H(A)$=17.55)]{%
   \includegraphics[width=.8\columnwidth]{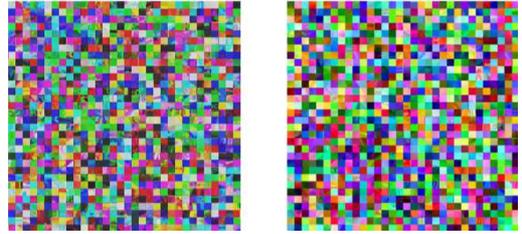}%
   \label{sfig:enc_prop}%
  }%

  \subfigure[Encrypted images using conventional scheme~\cite{IEICE-T2017:KKurihara} \quad \quad (Airplane: $H(A)$=15.62, Tiffany: $H(A)$=16.66)]{%
   \includegraphics[width=.8\columnwidth]{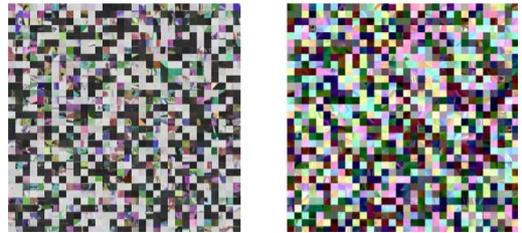}%
   \label{sfig:enc_conv}%
  }%
 \caption{Encrypted images using proposed and conventional
  schemes (Airplane and Tiffany).}
 \label{fig:encrypted_images}
 \end{figure}

 \begin{table}[t]
  \begin{center}
  \caption{Entropies of encrypted images using proposed and conventional schemes and original images.}
  \label{tb:entropy}
   \begin{tabular}{|c|c|c|c|} \hline
      & Prop. & Conv. & Original \\ \hline\hline
    Airplane & 17.76 & 15.62 & 13.96 \\ \hline
    Tiffany & 17.55 & 16.66 & 14.66 \\ \hline
    Lena & 17.96 & 17.74 & 16.84 \\ \hline
    Mandrill & 17.97 & 17.90 & 17.74 \\ \hline
    Peppers & 17.94 & 17.63 & 17.03 \\ \hline
    Sailboat & 17.96 & 17.62 & 16.88 \\ \hline
    Splash & 17.73 & 17.16 & 15.52 \\ \hline
   \end{tabular}
  \end{center}
 \end{table}

 \begin{figure*}[t]
 \centering

  \subfigure[Proposed]{%
    \includegraphics[width=.6\columnwidth]{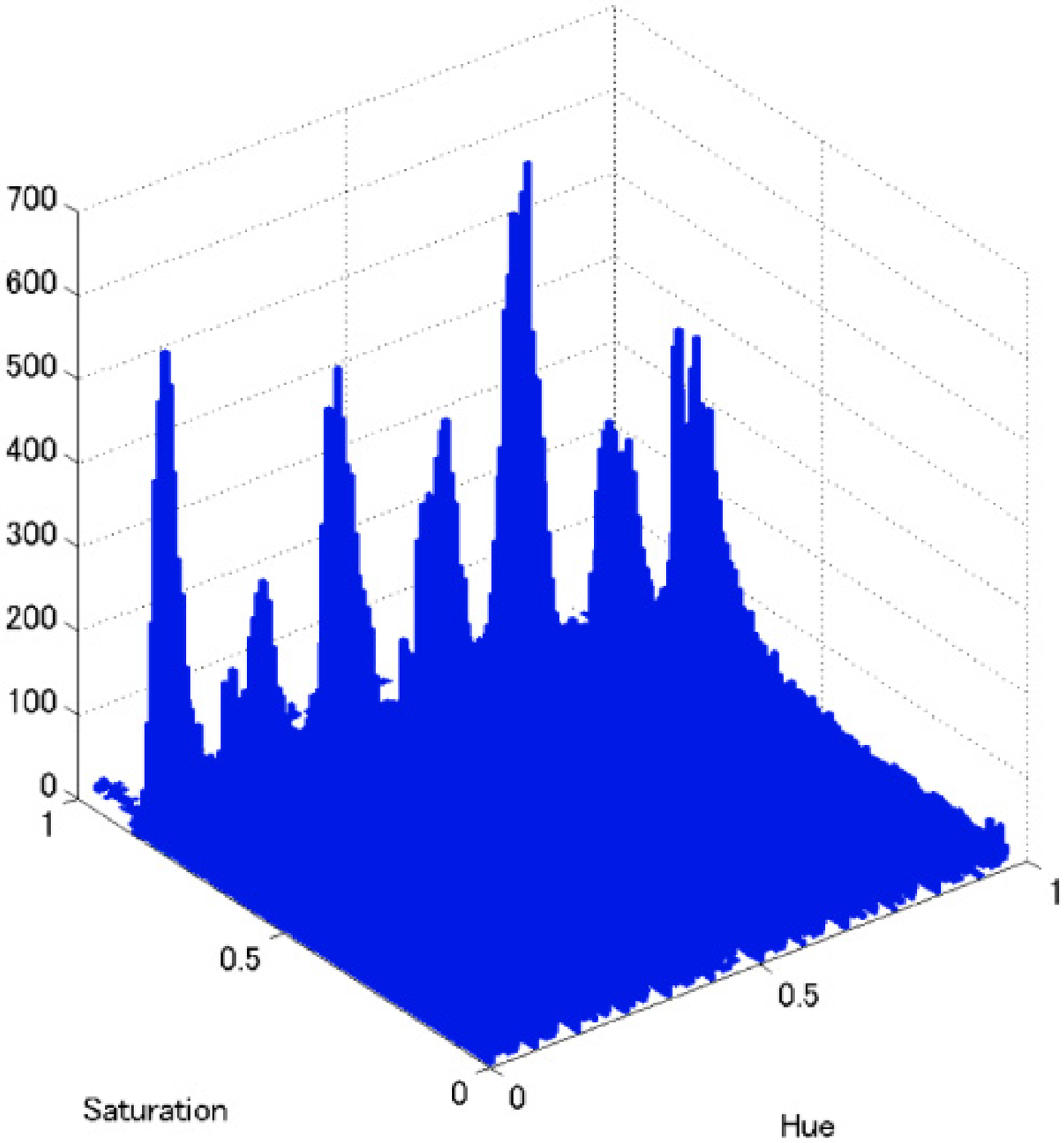}%
    \label{sfig:prophist_air}%
   }%
   \hfil%
  \subfigure[Conventional]{%
   \includegraphics[width=.6\columnwidth]{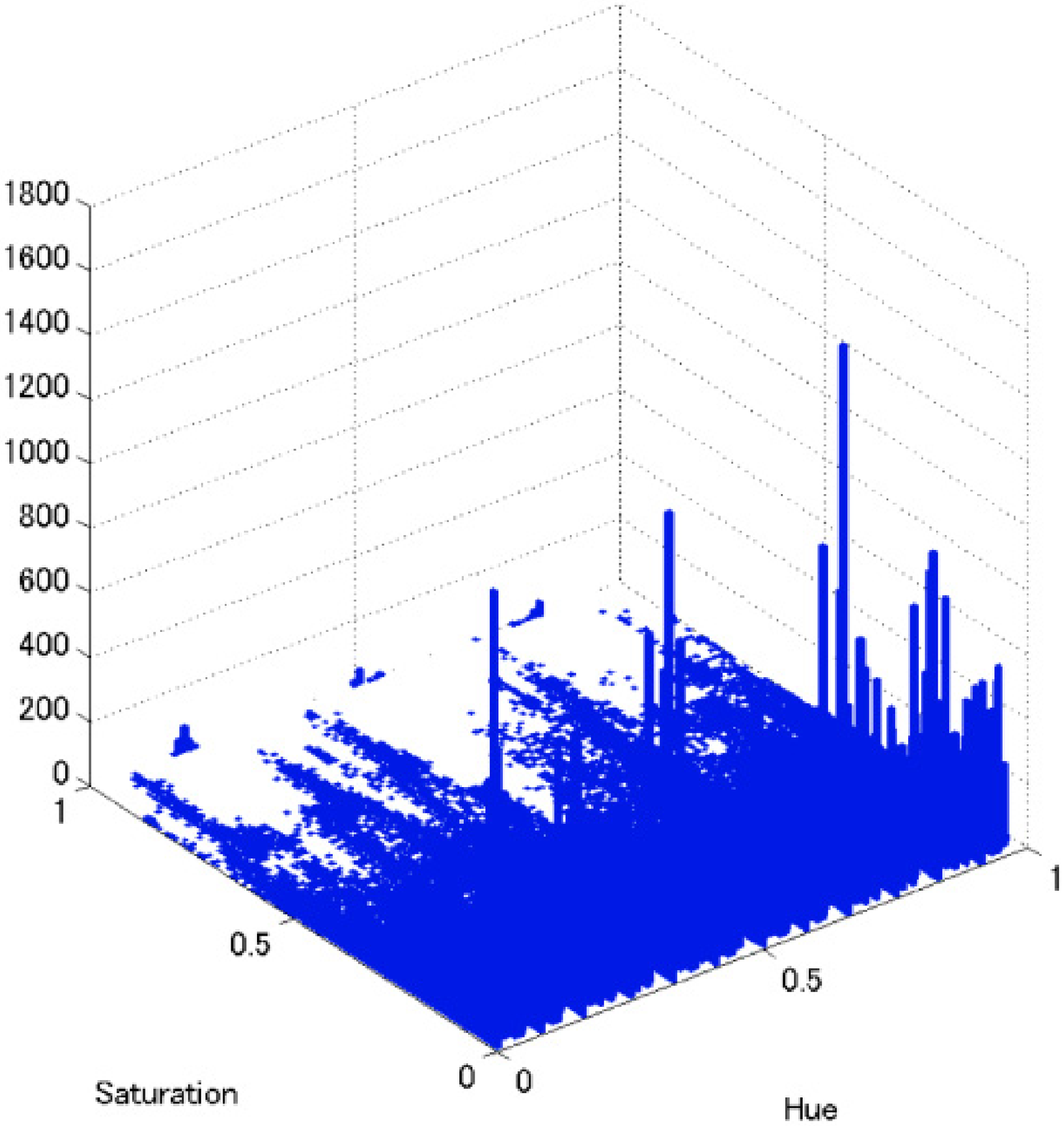}%
   \label{sfig:convhist_air}%
  }%
  \hfil%
  \subfigure[Original]{%
   \includegraphics[width=.6\columnwidth]{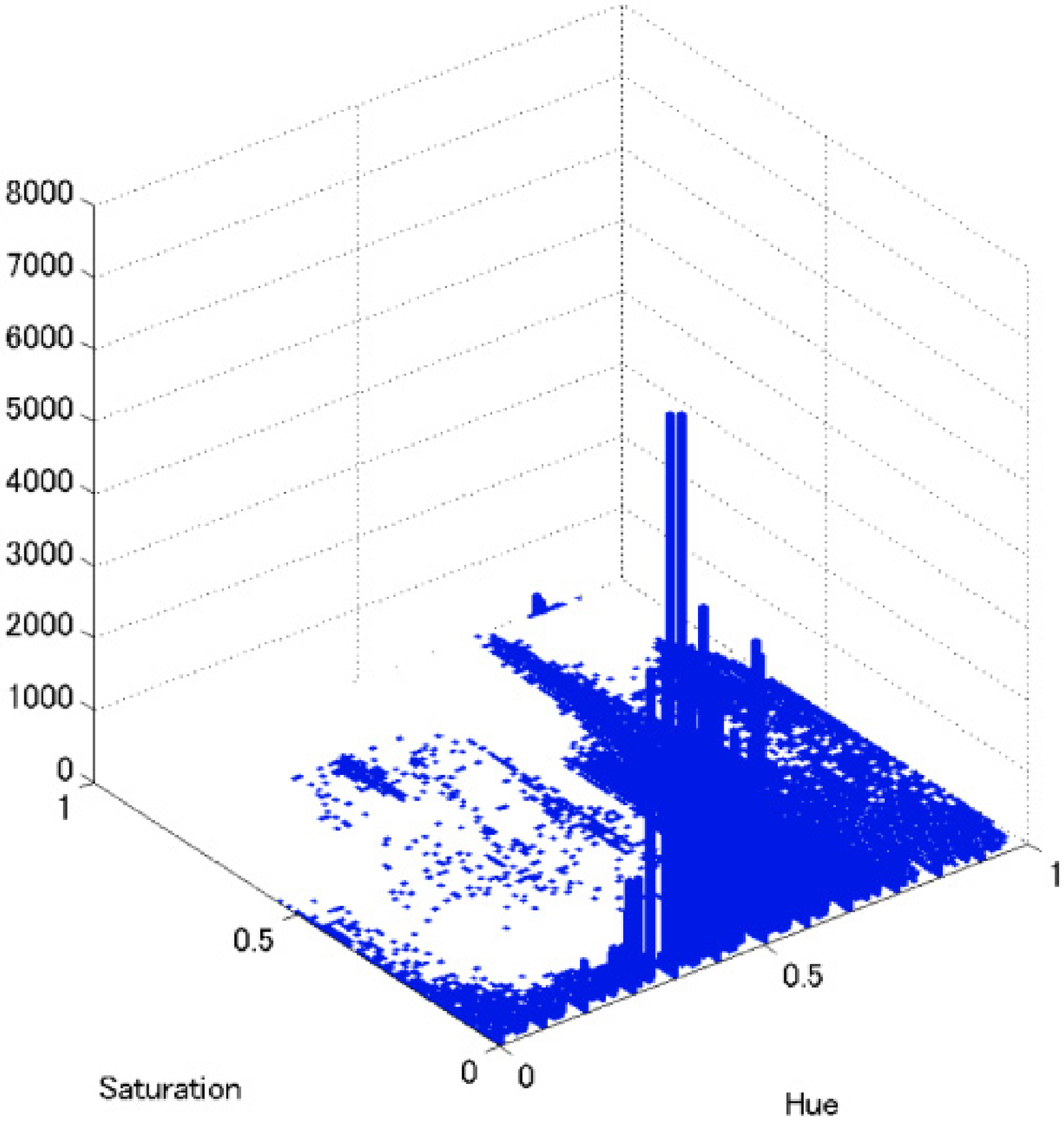}%
   \label{sfig:orighist_air}%
  }%
 \caption{Color distribution (Airplane).}
 \label{fig:3Dhistgram_air}
 \end{figure*}

 \begin{figure*}[t]
 \centering
 
   \subfigure[Proposed]{%
    \includegraphics[width=.6\columnwidth]{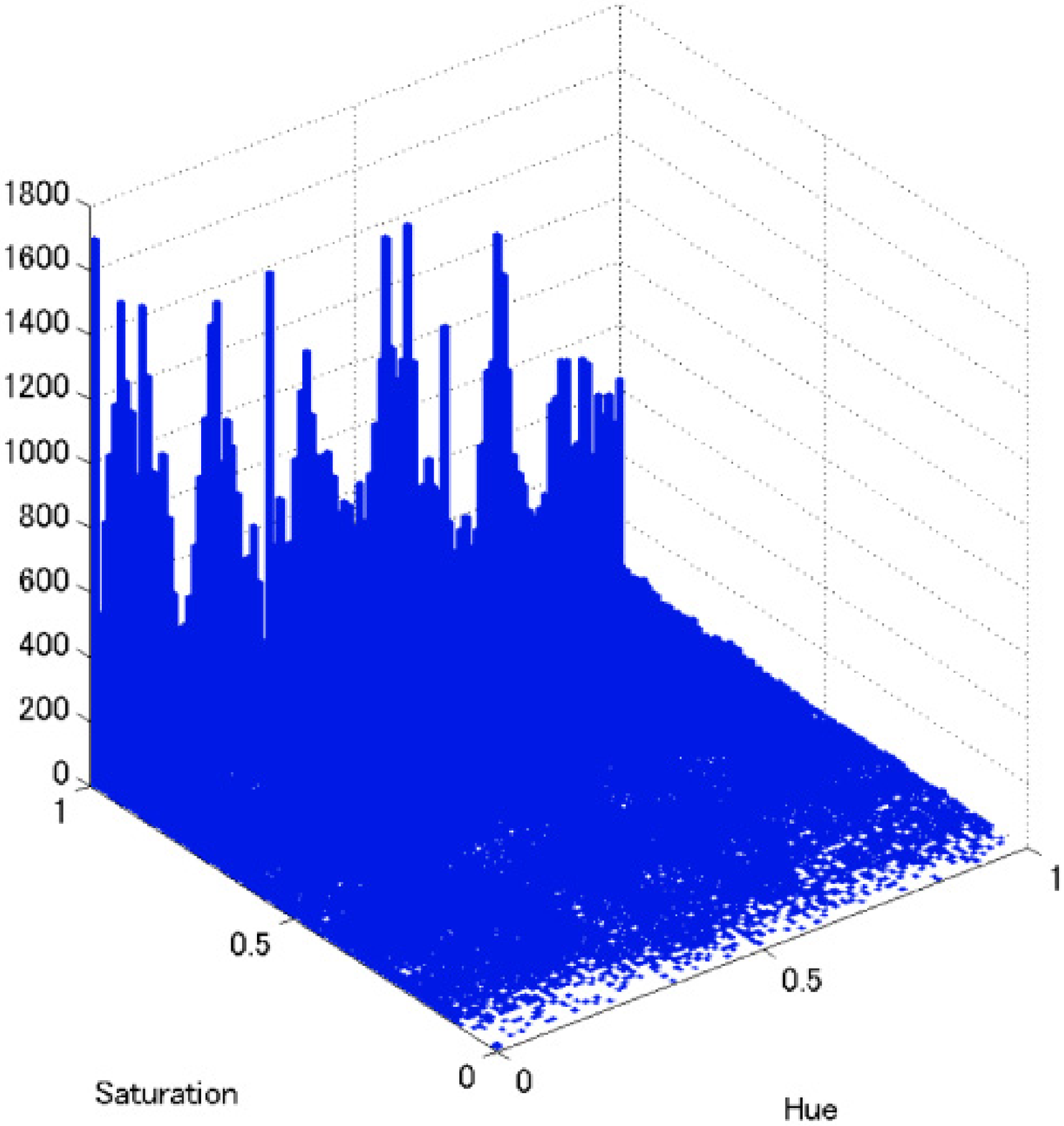}%
    \label{sfig:prophist_tif}%
   }%
   \hfil%
  \subfigure[Conventional]{%
   \includegraphics[width=.6\columnwidth]{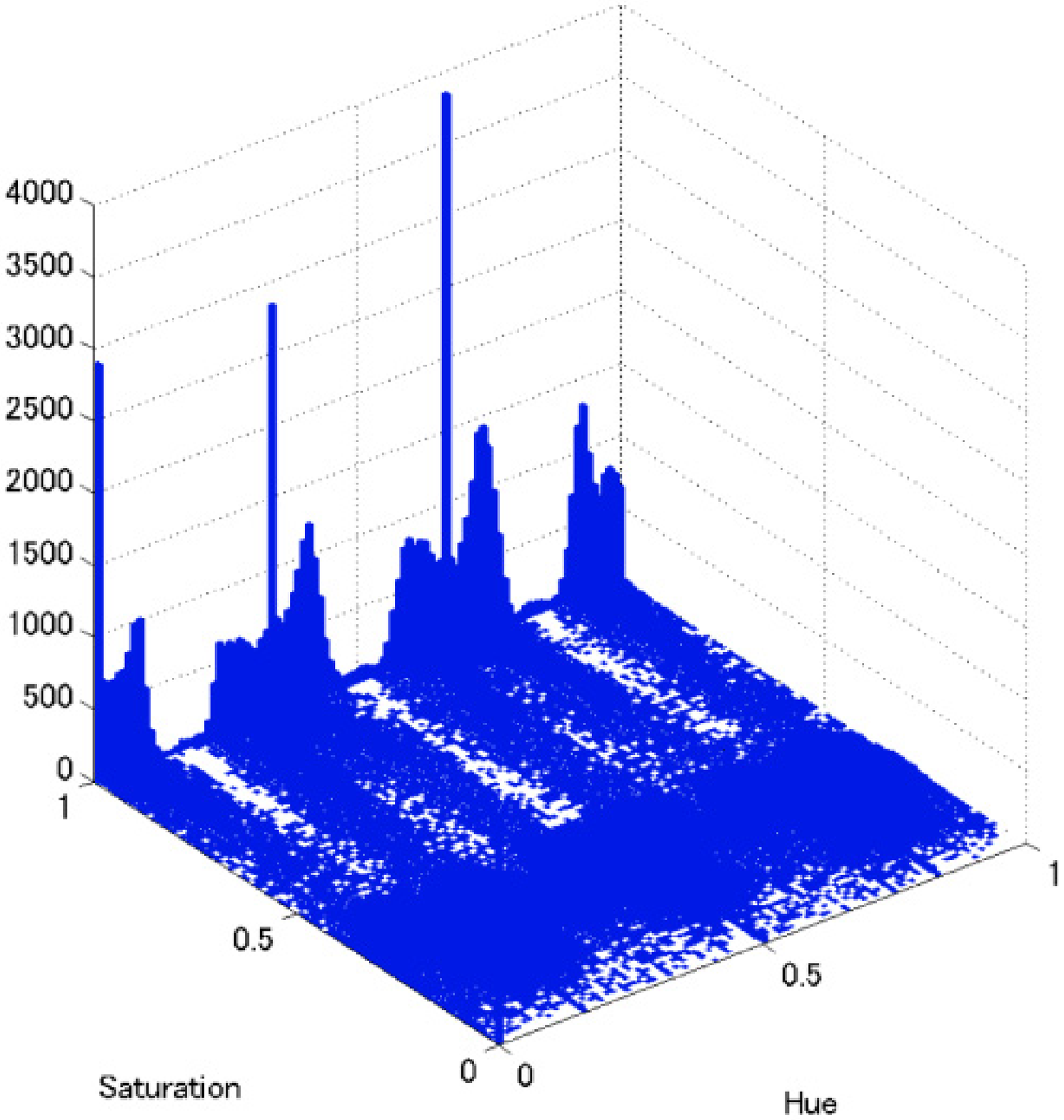}%
   \label{sfig:convhist_tif}%
  }%
  \hfil%
  \subfigure[Original]{%
   \includegraphics[width=.6\columnwidth]{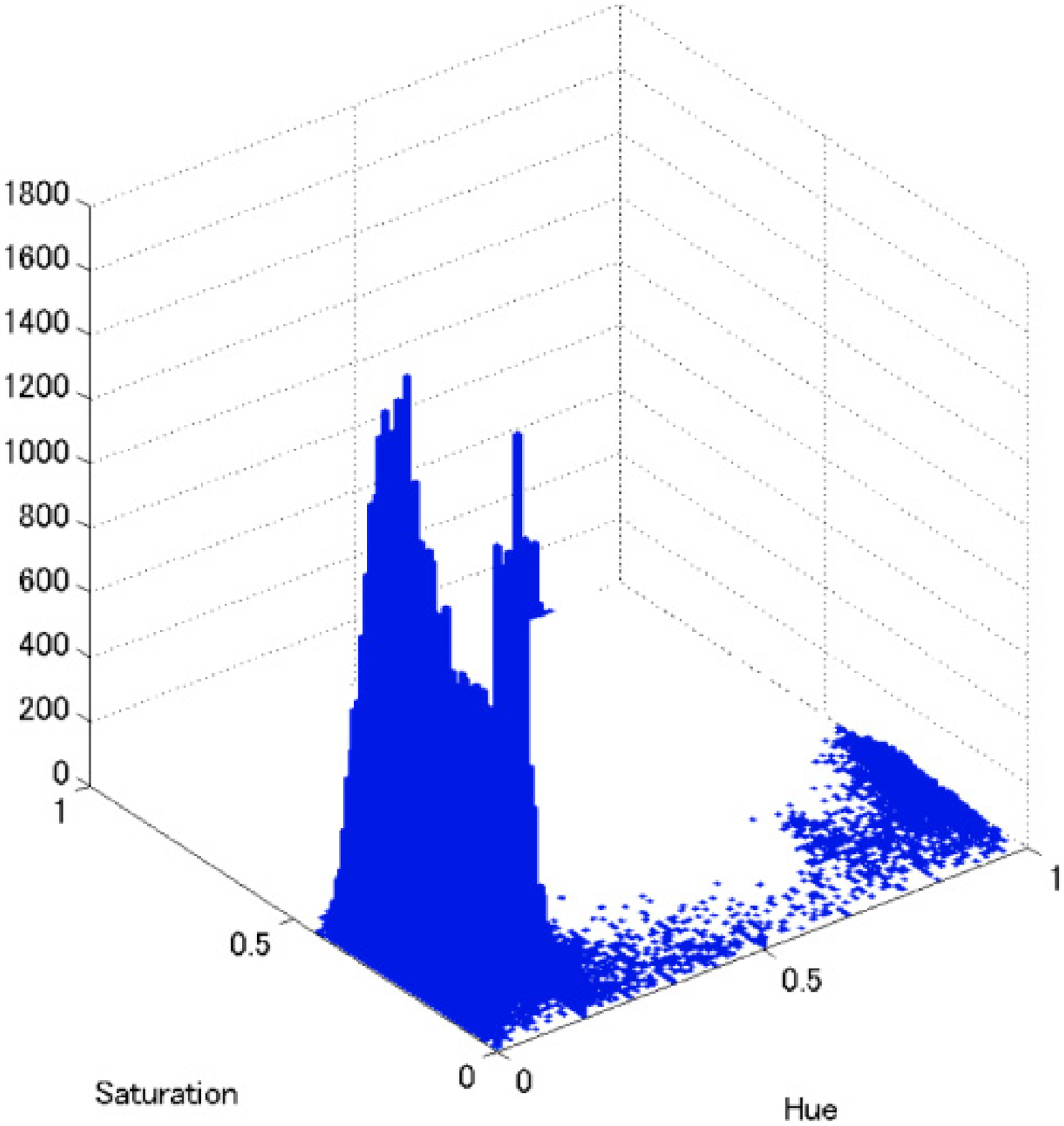}%
   \label{sfig:orighist_tif}%
  }%
 \caption{Color distribution (Tiffany).}
 \label{fig:3Dhistgram_tif}
 \end{figure*}

 Fig.~\ref{fig:encrypted_images} shows two of the original test images and their encrypted images produced by the proposed and the conventional schemes, where the divided block size is $16 \times 16$ pixels, respectively. 
 Table~\ref{tb:entropy} indicates the entropies of the original images and their encrypted images by using the proposed and the conventional schemes. 
 Entropy $H(A)$ is defined as
 \begin{eqnarray}
  H(A) = -\sum^{2^{24}}_{i=1}p(a_i)\log_{2}p(a_i), \label{eq:H}
 \end{eqnarray}
 where $A$ represents the finite set of 24-bit color $a_i~(i = 1, 2, ..., 2^{24})$, that is, $A = \{a_1, a_2, ..., a_{2^{24}}\}$, and $p(a_i)$ is the occurrence probability of $a_i$. 
 The encrypted images produced by the proposed scheme have obtained higher entropies than those of both the original images and the encrypted images produced by the conventional scheme.

 In addition, we compare the color distributions among the encrypted images by using the proposed and the conventional schemes and the original image. 
 Figs.~\ref{fig:3Dhistgram_air} and \ref{fig:3Dhistgram_tif} show the histograms of the test images shown in Fig.~\ref{fig:encrypted_images}, where the vertical/horizontal axes represent the saturation/hue values, respectively. 
 The encrypted images produced by the proposed scheme show wider distributions relative to the encrypted images produced by the conventional scheme and their original images.

\subsection{Compression efficiency}
\label{ssec:4-3}
 \begin{figure*}[t]
  \centering

  \subfigure[Airplane]{%
    \includegraphics[width=.95\columnwidth]{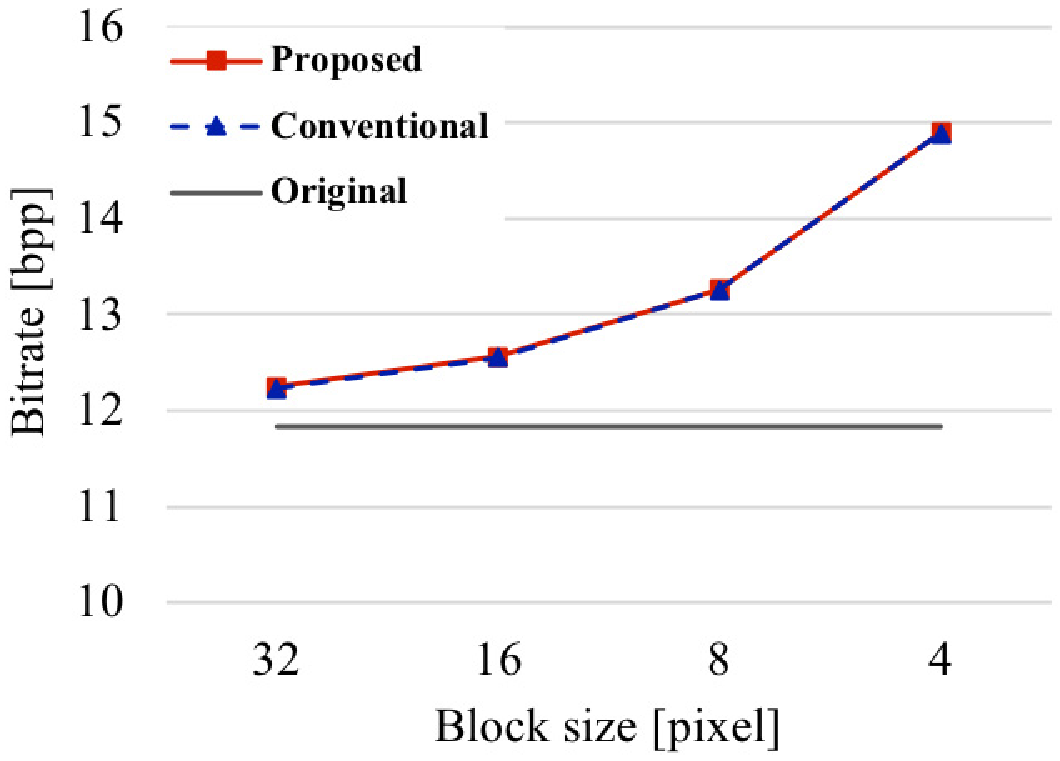}%
    \label{sfig:ce_air}%
   }%
   \hfil%
  \subfigure[Tiffany]{%
   \includegraphics[width=.95\columnwidth]{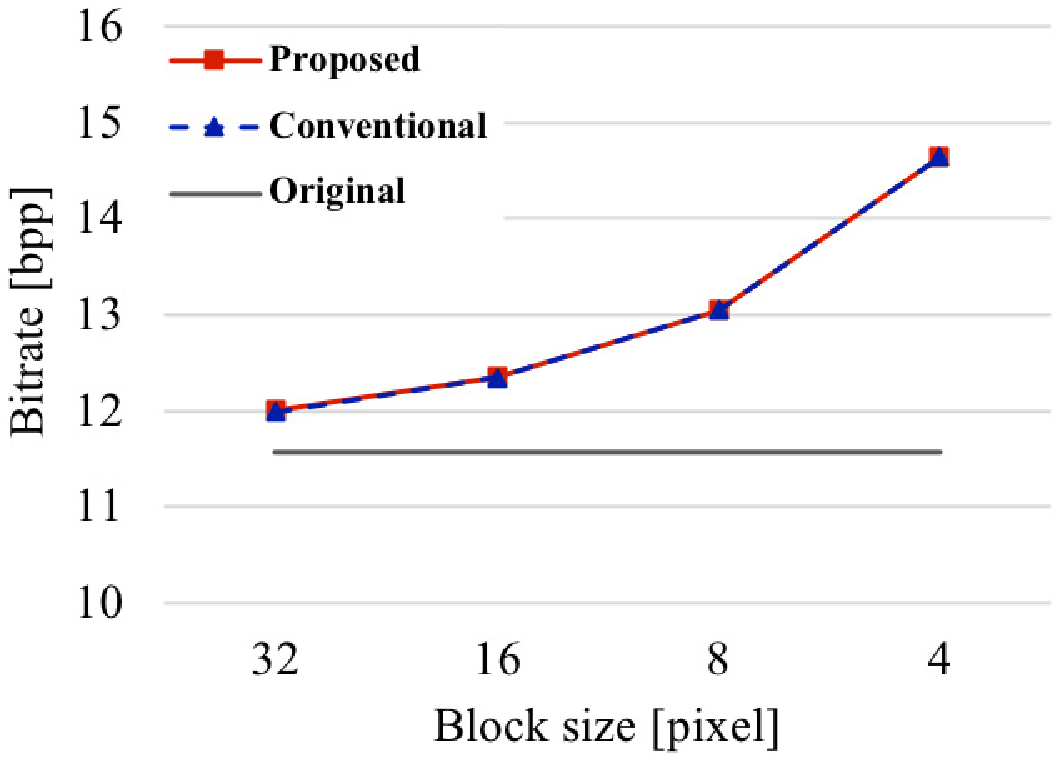}%
   \label{sfig:ce_tif}%
  }%
   \caption{Compression efficiency using JPEG-LS.}
   \label{fig:ce}
 \end{figure*}

 The lossless compression using JPEG-LS~\cite{IEEE-T:MJWeinberger} is performed on the encrypted images obtained by the proposed and the conventional schemes. 
 The compression efficiency is compared by calculating the bitrates.
 Fig.~\ref{fig:ce} shows the comparative results for the two encrypted images shown in Fig.~\ref{fig:encrypted_images}. 
 The sizes of the divided blocks are $4 \times 4$, $8 \times 8$, $16 \times 16$, and $32 \times 32$ pixels in Fig.~\ref{fig:ce}. 
 The results for the other five test images have analogous lines. 
 It is shown that the proposed scheme can maintain almost the same compression efficiency as the conventional scheme.
 
 Image compression generally utilizes the correlation among the pixels and/or the correlation within each block rather than that over the whole image. 
 A BPBE image can be obtained through four processes: positional scrambling, block rotation/flip, negative-positive transformation, and color component shuffling. 
 Therefore, the proposed BPBE scheme, which processes the RGB components independently, can maintain the correlation among the blocks, and thus does not severely degrade the compression efficiency compared to the conventional scheme, which processes the RGB components commonly. 
 Additionally, our scheme assigns different keys to each color component and independently processes those color components to increase the complexity among the blocks. 
 It is necessary for JPS to estimate the relationship over the whole image to retrieve the original image. 
 The proposed scheme can consequently increase the resilience against JPS.

\section{Conclusions}
\label{sec:5}
 We proposed a BPBE algorithm to increase the extent of the scrambling in this paper.
 The proposed scheme takes into account the RGB-based compression, such as JPEG-LS.
 Therefore, our scheme can scramble the RGB components independently by assigning different keys to each color component without degrading the JPEG-LS compression efficiency.
 The color distribution of the encrypted image produced by our scheme is consequently expanded compared to that of the conventional scheme.
 The resilience against JPS is simultaneously enhanced in the proposed scheme.
 Furthermore, the proposed scheme has exponentially extended the key space for resilience against brute-force attacks. 

\section*{Acknowledgements}
The authors would like to thank T. Ogasawara for sharing his idea that was treated in this work.
We also thank T. Chuman for collaborating in the experimental stages.
This work was partially supported by Grant-in-Aid for Scientific Research(B), No.17H03267, from the Japan Society for the Promotion of Science.





\begin{thebibliography}{99}
 \bibitem{Elsevier_DSP2017:MKumar} M. Kumar and A. Vaish, ``An Efficient 
 Encryption-then-Compression Technique for Encrypted Images Using SVD,'' 
 Digital Signal Processing, vol.60, pp.81--89, 2017.
 \bibitem{IEEE-T2014:JZhou} J. Zhou, X. Liu, O.C. Au, and Y.Y. Tang,
	 ``Designing an Efficient Image Encryption-Then-Compression
	 System via Prediction Error Clustering and Random
	 Permutation,'' IEEE Trans. Information Forensics and Security,
	 vol.9, no.1, pp.39-50, 2014.
 \bibitem{IEEE-T2010:WLiu} W. Liu, W. Zeng, L. Dong, and Q. Yao,
	 ``Efficient Compression of Encrypted Gray-Scale Images,'' IEEE
	 Trans. Image Process., vol.19, no.4, pp.1097-1102, 2010.
 \bibitem{IEEE-T2004:MJohnson} M. Johnson, P. Ishwar, V. Prabhakaran,
	 D. Schinberg, and K. Ramchandran, ``On Compressing Encrypted
	 Data,'' IEEE Trans. Signal Process., vol.52, no.10,
	 pp.2992-3006, 2004. 
 \bibitem{IEICE-T2017:KKurihara} K. Kurihara, S. Imaizumi, S. Shiota,
	 and H. Kiya, ``An Encryption-then-Compression System for
	 Lossless Image Compression Standards,'' IEICE Trans.  Inf. \&
	 Sys., vol.E100-D, no.1, pp.52-56, 2017.
 \bibitem{BMSB2016:KKurihara} K. Kurihara, O. Watanabe, and H. Kiya, 
 ``An Encryption-then-Compression System for JPEG XR Standard,'' in 
 Proc.on IEEE International Symposium on Broadband Multimedia Systems 
 and Broadcasting (BMSB), pp.1-5, 2016.
 \bibitem{IEICE-T2015:KKurihara} K. Kurihara, M. Kikuchi, S. Imaizumi,
	 S. Shiota, and H. Kiya, ``An Encryption-then-Compression System
	 for JPEG/Motion JPEG Standard,'' IEICE Trans. Fundamentals,
	 vol.E98-A, no.11, pp.2238-2245, 2015.
 \bibitem{ICASSP2015:OWatanabe} O. Watanabe, A. Uchida, T. Fukuhara, and
	 H. Kiya, ``An Encryption-then-Compression System for JPEG 2000
	 Standard,'' in Proc. on IEEE ICASSP, pp.1226-1230, 2015.
 \bibitem{GPEM2016:DSholomon} D. Sholomon, O.E. David, and
	 N.S. Netanyahu, ``An Automatic Solver for Very Large Jigsaw
	 Puzzles using Genetic Algorithms,'' Genetic Programming and
	 Evolvable Machines, vol.17, no.3, pp.291-313, 2016.
 \bibitem{CVPR2012:ACGallagher} A.C. Gallagher, ``Jigsaw Puzzles with
	 Pieces of Unknown Orientation,'' in Proc. on CVPR, pp.382-389,
	 2012.
 \bibitem{CVPR2016:KSon} K. Son, D. Moremo, J. Hays, and
	 D.B. Cooper, ``Solving Small-Piece Jigsaw Puzzles by Growing
	 Consensus,'' in Proc. on CVPR, pp.1193-1201, 2016.
 \bibitem{CVPR2015:GPaikin} G. Paikin and A. Tal, ``Solving Multiple
	 Square Jigsaw Puzzles with Missing Pieces,'' in Proc. on CVPR,
	 pp.4832-4839, 2015.
 \bibitem{ECCV2014:KSon} K. Son, J. Hays, and D.B. Cooper, ``Solving 
 Square Jigsaw Puzzles with Loop Constraints,'' in Proc. on ECCV, 
 vol.8694, pp.32-46, 2014.
 \bibitem{CVPR2011:DPomeraz} D. Pomeranz, M. Shemesh, and O. Ben-Shahar, 
 ``A Fully Automated Greedy Square Jigsaw Puzzle Solver,'' in Proc. on 
 CVPR, pp.9-16, 2011.
 \bibitem{CVPR2010:TCho} T.S. Cho, S. Avidan, and W.T. Freeman, ``A 
 Probabilistic Image Jigsaw Puzzle Solver,'' in Proc. on 
 CVPR, pp.183-190, 2010.
 \bibitem{IEEE-T:MJWeinberger} M.J. Weinberger, G. Seroussi, and
	 G. Sapiro, ``The LOCO-I Lossless Image Compression Algorithm:
	 Principles and Standardization into JPEG-LS,'' IEEE
	 Trans. Image Process., vol.9, no.8, pp.1309-1324, 2000.
 \bibitem{IEICE-T2018:TChuman} T. Chuman, K. Kurihara, and H. Kiya, ``On 
 the Security of Block Scrambling-Based EtC Systems against 
 Extended Jigsaw Puzzle Solver Attacks,'' IEICE Trans.  Inf. \&
	 Sys., vol.E101-D, no.1, pp.37-44, 2018.
 \bibitem{ICASSP2017:TChuman} T. Chuman, K. Kurihara, and H. Kiya, ``On
	 the Security of Block Scrambling-based ETC Systems against
	 Jigsaw Puzzle Solver Attacks,'' in Proc. of ICASSP, pp.2157-2161, 2017.
 \bibitem{ICME2017:TChuman}  T. Chuman, K. Kurihara, and H. Kiya, 
 ``Security evaluation for block scrambling-based etc systems against 
 extended jigsaw puzzle solver attacks,'' in Proc. on ICME, pp.229-234, 2017.
 \bibitem{ICME2018:WSirichotedumrong} W. Sirichotedumrong, T. Chuman, S. Imaizumi, and H. Kiya, 
 ``Grayscale-Based Block Scrambling Image Encryption for Social Networking Services,'' 
 in Proc. on ICME, 2018.
 \bibitem{arXiv:RYu} R. Yu, C. Russell, and L. Agapito, ``Solving Jigsaw Puzzles with Linear Programming,'' 
 arXiv preprint arXiv:1511.04472, 2015.
\end{thebibliography}
%
\profile{Shoko IMAIZUMI}{received her B.Eng., M.Eng., and Ph.D. degrees from Tokyo Metropolitan University, Japan in 2002, 2005, and 2011, respectively. 
In 2011, she joined Chiba University, where she is currently an Associate Professor of Graduate School of Engineering. 
From 2003 to 2004, she was with the Ministry of Education, Culture, Sports, Science and Technology of Japan. 
She was a Researcher at the Industrial Research Institute of Niigata Prefecture from 2005 to 2011. 
Her research interests include image processing and multimedia security. 
Dr. Imaizumi serves as an Associate Editor for IEICE Trans. Fundamentals and a Director for SPIJ (Society of Photography and Imaging of Japan). 
She is a member of IEICE, ITE, SPIJ, APSIPA, and IEEE.}
\profile{Hitoshi KIYA}{received his B.E and M.E. degrees from Nagaoka University of Technology, in 1980 and 1982 respectively, and his Dr. Eng. degree from Tokyo Metropolitan University in 1987. In 1982, he joined Tokyo Metropolitan University, where he became Full Professor in 2000. From 1995 to 1996, he attended the University of Sydney, Australia as a Visiting Fellow. He is a Fellow of IEEE, IEICE and ITE. He currently serves as President-Elect of APSIPA, and he served as Inaugural Vice President (Technical Activities) of APSIPA in 2009-2013, and as Regional Director-at-Large for Region 10 of IEEE Signal Processing Society in 2016-2017. He was also President of IEICE Engineering Sciences Society in 2011-2012, and he served there as Vice President and Editor-in-Chief for IEICE Society Magazine and Society Publications. He was Editorial Board Member of eight journals, including IEEE Trans. on Signal Processing, Image Processing, and Information Forensics and Security, Chair of two technical committees and Member of nine technical committees including APSIPA Image, Video, and Multimedia Technical Committee (TC), and IEEE Information Forensics and Security TC. He has organized a lot of international conferences, in such roles as TPC Chair of IEEE ICASSP 2012 and General Co-Chair of IEEE ISCAS 2019. Dr. Kiya was a recipient of numerous awards, including six best paper awards.}

%
\end{document}